\documentclass[a4paper,11pt, english,notitlepage,sumlimits]{article}

\usepackage[english]{babel}
\usepackage[utf8x]{inputenc}
\usepackage[a4paper,top=3cm,bottom=3cm,left=3cm,right=3cm,marginparwidth=1.75cm]{geometry}
\usepackage{dsfont}
\usepackage{mathtools} 
\usepackage{subfig}
\usepackage{multirow}
\usepackage{xcolor}
\usepackage{enumitem}
\usepackage{amssymb}
\usepackage{fancyhdr}
\usepackage{lmodern}
\usepackage[T1]{fontenc}
\usepackage{caption}
\usepackage{float}
\usepackage{cancel} 
\usepackage{graphics} 
\usepackage{amsthm}
\usepackage{revsymb}
\usepackage{authblk}
\usepackage{amsmath}

\usepackage[colorlinks=true,allcolors=black]{hyperref} 

\newtheorem{theorem}{Theorem}

\newtheorem{corollary}[theorem]{Corollary}

\newtheorem{lemma}[theorem]{Lemma}

\newlength{\blank}
\newenvironment{myproof}[1][{\hspace{-\blank}}]{{\medskip\noindent\textbf{Proof~{#1}.\ }}}{\hfill$\blacksquare$}

\newcommand{\tr}{\operatorname{Tr}}
\newcommand{\1}{\openone}
\newcommand{\id}{\operatorname{id}}

\newcommand{\ket}[1]{\vert #1\rangle}
\newcommand{\bra}[1]{\langle #1\vert}
\newcommand{\proj}[1]{\ket{#1}\!\bra{#1}}

\begin{document}

\title{Certifying bipartite entangled states \protect\\ 
with few local measurements: \protect\\
from separable stabilizers to applications}

\author[1,a]{Jennifer Ahiable}
\author[1,2,3,4,b]{Andreas Winter}

\affil[1]{{\small\it Grup d'Informaci\'o Qu\`antica, Departament de F\'isica, \protect\\ Universitat Aut\`onoma de Barcelona, 08193 Bellaterra (Barcelona), Spain\vspace{1mm}}}
\affil[2]{{\small\it ICREA---Instituci\'o Catalana de Recerca i Estudis Avan\c{c}ats, \protect\\ Pg.~Lluis Companys, 08010 Barcelona, Spain\vspace{1mm}}}
\affil[3]{{\small\it Department Mathematik/Informatik---Abteilung Informatik,\protect\\ Universit\"at zu K\"oln, Albertus-Magnus-Platz, 50923 K\"oln, Germany\vspace{1mm}}}
\affil[4]{{\small\it Institute for Advanced Study, Technische Universit\"at M\"unchen,\protect\\ Lichtenbergstra{\ss}e 2a, 85748 Garching, Germany\vspace{1mm}}}
\affil[a]{{\small\textit{Email:} jennifer.ahiable@uab.cat}}
\affil[b]{{\small\textit{Email:} andreas.winter@uni-koeln.de}}

\date{4 December 2025}

\maketitle

\begin{abstract}
We show a simple and systematic way to certify any given bipartite state as the unique joint $1$-eigenstate of two separable projectors, each of which can be measured with simple local observables. This is practically useful, as the detection probabilities of the two stabilizer projectors relate directly to the fidelity of certification. The same result gives a simple and effective lower bound on the entanglement fidelity of a quantum channel in terms of two ensemble fidelities. 

We then generalise the bipartite result recursively to multipartite systems, showing that every $n$-party pure state is the unique joint $1$-eigenstate of $2^{n-1}$ separable projectors, and an upper bound of the infidelity of the state in terms of the infidelities of the separable stabilizer projectors. 
\end{abstract}

\section{Introduction}
Entanglement is not only the crucial feature of quantum mechanics that marks its departure from classical thinking, it is the fuel of quantum information processing and quantum technologies \cite{Bennett:Tele, Ekert1991,Nielsen}. Because of this, being able to certify by suitable measurements that a given experiment has prepared a desired entangled state is a core concern in many applications. The certification should yield an estimation of the quality of the actual state with respect to the ideal target in terms of measured quantities, ideally with easily controlled error bars.
What precisely this certification entails is not a priori defined, but given the distributed nature of the entangled systems, one common requirement will be that the measurements involved be ``local''. Formally, a state $\rho$ is called separable \cite{Werner} if it can be written as a classical mixture of product states, $\rho = \sum_k \; p_k \rho^{(k)}_1 \otimes \cdots \otimes \rho^{(k)}_n$, with probabilities satisfying $p_k \geq 0,$ and $\sum_k p_k =1$. For $n=2$, this defines bipartite separability, while for $n > 2$, we mean the state is fully separable; otherwise it is entangled \cite{Horodeckix4:review}. Local measurements are described by separable operators, which are those of the form $M = \sum_k M^{(k)}_1 \otimes \cdots \otimes M^{(k)}_n$ and are implementable with local operations and classical communication (LOCC) \cite{LOCC, Nielsen}. Under this constraint, certification may range from full density matrix tomography (for an arbitrary mixed target state) \cite{BertlandKrammer:fullTom, Artiles}, to shadow tomography \cite{Aaronson} of the fidelity (for a pure target state) \cite{HuangPreskillSoleimanifar}, to specially designed experimentally lean procedures for certain states \cite{Bavaresco:PhD,Ohad-et-al}. 

Recently, it was realised and experimentally demonstrated \cite{Bavaresco-et-al:fid,YuShangGuehne-1,YuShangGuehne-2,Huang-et-al,Ohad-et-al} that arbitrary-dimensional pure entangled states can be efficiently certified by a few local measurements in a bipartite system, and similarly for certain multipartite states. 

\medskip
In the present work, we show a simple variant of the method in \cite{Bavaresco-et-al:fid}, which lends itself directly to a realisation of the test as one-way LOCC \emph{(and which had been described earlier in \cite{YuShangGuehne-1})}. The key to our approach is identifying any target bipartite pure state $\ket{\psi}$ as the unique joint +1-eigenstate of two suitable separable projectors. The first projector performs a correlation check in the state's natural Schmidt basis, while the second performs a complementary check in a conjugate basis. Employing this technique, we show that measurements of the two projectors allow for a direct and experimentally robust lower bound on the fidelity $F\left(\rho,\psi\right)$\footnote{Here and throughout, we denote a pure state vector $\ket{\psi}$, while its projector is $\psi \equiv \proj{\psi}$.} of the prepared state $\rho$ to the target state $\ket{\psi}$. 
We extend the utility of this framework beyond state certification, demonstrating that the same result provides an effective lower bound on the entanglement fidelity of a quantum channel, which can be estimated by testing the channel on pure states.

Finally, we establish the scalability of our approach by generalizing the entire method recursively to the multipartite setting. The procedure can be understood as a sequential cascade of bipartite certifications. We begin by treating the system as a bipartite split between the first party and the remaining $n-1$ parties. After the first party performs their measurements and communicates the results, the process is repeated for the second party and the rest, continuing sequentially down the line. This iterative method results in a set of $2^{n-1}$ fully separable projectors that collectively act as a unique stabilizer for the target $n$-party state. Crucially, this generalization preserves the experimental robustness of the original method, yielding a direct lower bound on the fidelity of the multipartite state from the measured outcomes of these simple, sequential tests. This provides a systematic and experimentally lean protocol for certifying any pure multipartite entangled state, in arbitrarily large local dimension.

The paper is structured as follows: Section \ref{sec:stabilizers} reviews the certification of stabilizer states in multi-qudit systems to motivate what we then do in general and to gain intuitions. 
Section \ref{sec:Core lemma} introduces our main result: the construction of two separable projectors that uniquely stabilize any bipartite pure state. Using this, we derive a robust fidelity bound and extend the method to certify quantum channels. In Section \ref{sec:multiparty}, we generalize this certification method to arbitrary multiparty pure states. Finally, we conclude in Section \ref{sec:open}  with a summary and a discussion of open questions.

\section{Stabilizer states certified by few local measurements}
\label{sec:stabilizers}
Let $\mathcal{P}_n$ be the $n$-qubit Pauli group, which consists of all $n$-fold tensor products of Pauli operators $\{\1,X,Y,Z\}$ with phases $\{\pm 1, \pm i\}$. For any $n$-qubit state $\ket{\psi}$, its stabilizer group $\mathcal{S} \subseteq \mathcal{P}_n$ is the set of all elements in the Pauli group that leave the state unchanged: 
\[
  \mathcal{S} = \{g\in\mathcal{P}_n: g\ket{\psi} = \ket{\psi}\}.
\]
The set $\mathcal{S}$ forms a maximal abelian subgroup of $\mathcal{P}_n$, and any state $\ket{\psi}$ defined by such a group is called a stabilizer state \cite{Gottesman:PhD,Hostens-et-al, NielsenAndChuang}. To uniquely specify a single $n$-qubit state, one must define $n$ independent and commuting generators $\{g_1,\ldots,g_n\}$ that form a basis for the group $\mathcal{S}$ \cite{Gottesman:PhD}. For instance, the Bell state $\ket{\phi^{+}} = \frac{1}{\sqrt{2}}(\ket{00} + \ket{11})$ is uniquely stabilized by the two unitary and Hermitian generators $g_1 = X \otimes X$ and $g_2 = Z \otimes Z$. The Pauli group and stabilizer formalism has been extended to higher-dimensional component systems (qudits), most notably to prime and prime power dimension \cite{AshikhminKnill,KKKS}, but the main features concerning us remain the same, so we will stick with qubits. 

The stabilizer properties can be used to certify the state in the most evident manner: the stabilizer group $\mathcal{S}$ and in particular the generators are automatically Hermitian, and hence up to sign tensor products of Paulis (which are themselves Hermitian with eigenvalues $\pm 1$). Thus, for each generator $g_k = P_{k,1}\otimes\cdots\otimes P_{k,n}$ one simply measures $P_{k,j}$ on the $j$-th qubit, obtaining a value $\pi_{k,j}\in\{\pm 1\}$, such that the measurement result of $g_k$ is the product $\prod_{j=1}^n \pi_{k,j}$. For each of the $g_k$ we have to verify that we obtain $+1$, ideally all the times the generator is measured. This is an entirely local procedure, requiring only the distribution of the information which generator $g_k$ is to be measured. Furthermore, each system is measured in only one of three possible bases. 

It is clear from the definition that any unique stabilizer state can also be described by the product of projectors formed from its generators. The operator $\Pi_k = \frac{1}{2}(\1 + g_k)$ is the projector onto the +1 eigenspace of the generator $g_k$. Since a unique stabilizer state $\ket{\psi}$ is the simultaneous $+1$-eigenstate of all its generators $g_1,\ldots,g_n$, the projector onto this state is simply the product of the individual projectors:
\begin{equation}
    \proj{\psi} = \prod_{k=1}^{n} \Pi_k =  \prod_{k=1}^{n}\frac{1}{2}(\1 + g_k).
\end{equation}
Note that, just as do the generators $g_k$, the projectors $\Pi_k$ commute pairwise. 
An important feature of this formalism is that the projectors $\Pi_k$ corresponding to stabilizer generators are themselves fully separable operators. This means that for any standard stabilizer state (which can of course be highly entangled), its identity can be confirmed by performing only a few local measurements, one for each generator. Each measurement of a generator $g_k$ on the state $\ket{\psi}$ will yield the outcome +1 with certainty, thus certifying the state.

Motivated by this powerful framework, our work seeks to generalize this principle beyond the Pauli group. We will show that any pure entangled state can be uniquely identified as the common +1 eigenstate of a small set of separable projectors, providing a universal method for certification with few local measurements.

\section{The core lemma}
\label{sec:Core lemma}
Given a bipartite pure state $\ket{\psi} \in A\otimes B$, we can write it in Schmidt basis form,
\begin{equation}
\label{eq:schmidt-dec}
  \ket{\psi} = \sum_{j=1}^d \sqrt{\lambda_j} \ket{j}^{A}\ket{j}^{B},
\end{equation}
with $\lambda_i\geq 0$ and $\sum_i \lambda_1=1$,
thus defining computational bases of $A$ and $B$, which we both may assume to be $d$-dimensional Hilbert spaces. 

Define a conjugate basis $\{\ket{\hat{\alpha}}^{A}\}$ of $A$ that is unbiased with respect to $\{\ket{j}^{A}\}$, i.e.
\begin{equation}
\label{eq:conj-basis}
  \ket{\hat{\alpha}}^{A} = \sum_{j=1}^d \frac{1}{\sqrt{d}} e^{i\varphi(j,\hat{\alpha})} \ket{j}^{A},
\end{equation}
with suitable phase angles $\varphi(j,\hat{\alpha})$. Depending on the dimension, there are usually many such bases \cite{TadejZyczkowski:Hadamard}, for example always the discrete Fourier basis, where $\varphi(j,\hat{\alpha}) = 2\pi \frac{j\hat{\alpha}}{d}$, or more generally when $j$ labels the elements of an abelian group $G$ with $d$ elements, the basis obtained by applying the Fourier transform of $G$ to the computational basis. 

Observe 
\begin{equation}
\label{eq:unit-vector}
 \bra{\hat{\alpha}}^{A}\ket{\psi} = \frac{1}{\sqrt{d}} \sum_{j=1}^d \sqrt{\lambda_j} e^{-i\varphi(j,\hat{\alpha})} \ket{j}^{B}
  =: \frac{1}{\sqrt{d}} \ket{\psi_{\hat{\alpha}}}^{B},
\end{equation}
where $\ket{\psi_{\hat{\alpha}}}^{B}$ is now a unit vector.

\begin{lemma}
  \label{lemma:little}
  Let $\ket{\psi} \in A\otimes B$ be an arbitrary pure state with Schmidt decomposition as seen in Eq.~\eqref{eq:schmidt-dec}. Let $\{\ket{\hat{\alpha}}^{A}\}$ be the conjugate basis of $A$ that is unbiased with respect to $\{\ket{j}^{A}\}$ as defined in Eq.~\eqref{eq:conj-basis}, and the relative unit state $\ket{\psi_{\hat{\alpha}}}^{B}$ on $B$ as in Eq.~\eqref{eq:unit-vector}. Define the operators$P=\sum_j \proj{j}^{A}\otimes\proj{j}^{B}$ and $Q=\sum_{\hat{\alpha}} \proj{\hat{\alpha}}^{A}\otimes\proj{\psi_{\hat{\alpha}}}^{B}$. Then, $P$ and $Q$ are projectors, and
  \begin{itemize}
      \item[(a)] $P\ket{\psi} = \ket{\psi} = Q\ket{\psi}$;
      \item[(b)] $PQ = QP = \proj{\psi}$.
  \end{itemize}
\end{lemma}
\begin{myproof}
The operators $P$ and $Q$ are projectors by definition. 
Point (a) is a straightforward calculation; (b) is not much more difficult:
\[\begin{split}
  PQ &= \sum_{j,\hat{\alpha}} \proj{j}^{A}\proj{\hat{\alpha}}^{A} \otimes \proj{j}^{B}\proj{\psi_{\hat{\alpha}}}^{B} \\
     &= \sum_{j,\hat{\alpha}} \ket{j}^{A} \frac{1}{\sqrt{d}} e^{i\varphi(j,\hat{\alpha})}  \bra{\hat{\alpha}}^{A}
        \otimes \ket{j}^{B} \sqrt{\lambda_j} e^{-i\varphi(j,\hat{\alpha})} \bra{\psi_{\hat{\alpha}}}^{B} \\
     &= \left( \sum_j \sqrt{\lambda_j}\ket{j}^{A}\ket{j}^{B} \right)
        \left( \sum_{\hat{\alpha}} \frac{1}{\sqrt{d}} \bra{\hat{\alpha}}^{A}\bra{\psi_{\hat{\alpha}}}^{B} \right)
      = \proj{\psi}, 
\end{split}\]
and similarly for $QP$. 
\end{myproof}

\medskip
This means that $\ket{\psi}$ is the unique joint $+1$-eigenstate of two projectors, which are both separable of rank $d$, so we have a stabilizer structure, familiar from Pauli stabilizers \cite{Gottesman:PhD,Hostens-et-al}: each projector eliminates a large part of the available Hilbert space, leaving a fraction $\frac{1}{d}$. 
What is more, not only $P$ and $Q$ are separable operators, but also their complementary projections $\1-P$ and $\1-Q$. In fact, the two measurements $(P,\1-P)$ and $(Q,\1-Q)$ can be implemented straightforwardly by a one-way LOCC procedure. For $P$, both parties measure in the computational basis $\{\ket{j}\}$, and accept if their outcomes $j_A$ and $j_B$ coincide, and reject otherwise. For $Q$, Alice measures in the conjugate basis $\{\ket{\hat{\alpha}}^{A}\}$, sends the result to Bob, who measures $\proj{\psi_{\hat{\alpha}}}^{B}$; he accepts if he gets outcome $1$ and rejects if he gets $0$. If it appears slightly inefficient that Bob only makes a binary measurement in this case, this can be improved by rescaling the projectors to effects $M_{\hat{\alpha}} := c_{\hat{\alpha}} \proj{\psi_{\hat{\alpha}}}^{B}$, with $0 < c_{\hat{\alpha}} < 1$ such that $0 \leq \sum_{\hat{\alpha}} M_{\hat{\alpha}} \leq \1$ (i.e.~they form part of a POVM).  
Compare \cite{Bavaresco-et-al:fid}, where a similar pair of measurements is employed. Not only does this lead to a characterisation of the state $\ket{\psi}$ as the unique state that passes both observables $P$ and $Q$ with outcome $+1$ with certainty, this characterisation is experimentally robust, as shown in the following theorem.

\begin{theorem}
  \label{theorem:fidelity}
  Given any state $\rho$ in the bipartite system $A \otimes B$, and $P$ and $Q$ as before,
  \[
    F\left(\rho,\psi\right)^2 := \tr\rho\psi \geq \tr\rho P + \tr\rho Q - 1. 
  \]
  Indeed, it even holds
  \begin{equation}
     \label{eq:fidelity-operator}
     \1-\proj{\psi} = \1-PQ \leq \1-P + \1-Q.
  \end{equation}
\end{theorem}
\begin{myproof}
It is clear that the fidelity inequality follows from Eq.~\eqref{eq:fidelity-operator}. The latter in turn follows from Lemma \ref{lemma:little} and the elementary relation $(\1-P)(\1-Q)\geq 0$, which is due to featuring a product of commuting positive semidefinite matrices on the left.
\end{myproof}

\medskip
This result can be interpreted as saying that if we measure $P$ and $Q$ and find, for example through sufficient repetition, that both have with high confidence and expectation $\tr \rho P, \tr \rho Q \geq 1-\epsilon$, then it follows that the fidelity of $\rho$ with our target pure state is $\geq 1-2\epsilon$. 

\medskip
We can apply this to the characterization of process fidelities, comparing the entanglement fidelity of a cptp map $\mathcal{T}$ acting on $B$ to its average fidelity with respect to judiciously chosen ensembles. The following corollary had been found previously by Hofmann \cite{Hofmann:fid} for the maximally entangled state.
In this respect, define for a pure state $\ket{\phi}^{RB}$,
\[
  F(\ket{\phi};\mathcal{T})^2 := \tr\proj{\phi}^{RB}\bigl((\id_R\otimes\mathcal{T})\proj{\phi}^{RB}\bigr),
\]
with an arbitrary reference system $R$. We shall use this notation both for trivial (i.e. one-dimensional) $R$, and for $R=A$. 

\begin{corollary}
  \label{cor:process-fidelity}
  Let $\mathcal{T}$ be a cptp map acting on $B$, and $\ket{\psi}\in A\otimes B$ be an arbitrary pure state. Then, with the above notation, 
  \begin{equation}
    \label{eq:process-general}
    F(\ket{\psi};\mathcal{T})^2 
    \geq \sum_j \lambda_j F(\ket{j}^{B};\mathcal{T})^2 
        + \sum_{\hat{\alpha}} \frac{1}{d} F(\ket{\psi_{\hat{\alpha}}}^{B};\mathcal{T})^2 
        -1.
  \end{equation}
\end{corollary}

\begin{myproof}
Define the state $\rho$ as the output of the channel $\mathcal{T}$ acting on subsystem $B$ of the pure state $\proj{\psi}$:
\begin{equation}
    \rho = (\id_A \otimes \mathcal{T})(\proj{\psi}).
\end{equation}
Following Theorem \ref{theorem:fidelity}, 
\begin{equation*}
    F(\ket{\psi}; \mathcal{T})^2 
     = F(\rho, \psi)^2 
     = \tr ((\id_A \otimes \mathcal{T})(\proj{\psi})\proj{\psi} 
     \geq \tr \rho P +  \tr \rho Q - 1.
\end{equation*} 
We can evaluate the two terms of the lower bound separately. For the first term,
\begin{align*}
  \tr \rho P 
   &= \tr((\id_A \otimes \mathcal{T})(\proj{\psi})) P \\
   &= \tr \left(\sum_{i,j} \sqrt{\lambda_i\lambda_j} \ket{i}\bra{j}^A \otimes \mathcal{T}(\ket{i}\bra{j}^B)\right)
   \!\!\left(\sum_k \proj{k}^A \otimes \proj{k}^B\right) \\
   &= \tr\left[ \sum_{i,j,k} \sqrt{\lambda_i\lambda_j} ( \ket{i}\bra{j}^A \proj{k}^A) \otimes (\mathcal{T}(\ket{i}\bra{j}^B) \proj{k}^B) \right] \\
   &= \tr\left[ \sum_{i,j,k} \sqrt{\lambda_i\lambda_k} ( \delta_{jk} \ket{i}\bra{k}^A) \otimes (\mathcal{T}(\ket{i}\bra{k}^B) \proj{k}^B) \right] \\
    &= \sum_{k} \lambda_k \tr \mathcal{T}(\proj{k}^{B}) \proj{k}^{B} \\
    &= \sum_k \lambda_k F(\ket{k}^{B}; \mathcal{T})^2.
\end{align*}

For the second term, we make use of the dual map $\mathcal{T}^*$:
\begin{align*}
  \tr \rho Q 
    &= \tr ((\id_A \otimes \mathcal{T})(\proj{\psi})) Q \\
    &= \tr \proj{\psi} (\id_A \otimes \mathcal{T}^*)(Q) \\
    &= \sum_{\hat{\alpha}} \tr \proj{\psi} \left(\proj{\hat{\alpha}}^A \otimes \mathcal{T}^*(\proj{\psi_{\hat{\alpha}}}^B) \right) \\
    &= \sum_{\hat{\alpha}} \bra{\psi} \left( \proj{\hat{\alpha}}^A \otimes \mathcal{T}^*(\proj{\psi_{\hat{\alpha}}}^B) \right)\ket{\psi} \\
    &= \sum_{\hat{\alpha}} \frac{1}{d} \bra{\psi_{\hat{\alpha}}}^B \mathcal{T}^*(\proj{\psi_{\hat{\alpha}}}^B) \ket{\psi_{\hat{\alpha}}}^B \\
    &= \sum_{\hat{\alpha}} \frac{1}{d} \tr \mathcal{T}(\proj{\psi_{\hat{\alpha}}}^{B}) \proj{\psi_{\hat{\alpha}}}^{B} \\
    &= \sum_{\hat{\alpha}} \frac{1}{d} F(\ket{\psi_{\hat{\alpha}}}^{B}; \mathcal{T})^2.
\end{align*}
Substituting these terms back into the inequality yields the desired result.
\end{myproof}

\medskip
In the same vein as the experiment-friendly entanglement certification before, this means that we can check that a process has high fidelity with the identity channel with respect to a certain pure state, without actually preparing an entangled state or making an entangled measurement, but simply by testing the channel on up to $2d$ pure states. Each term $F(\ket{\phi}^B;\mathcal{T})^2$ in turn is the expectation of the observable $\proj{\phi}^B$ on the state $\mathcal{T}(\proj{\phi^B})$, or equivalently the probability of acceptance of the binary POVM $(\proj{\phi}^B,\1-\proj{\phi}^B)$.

However, we do not need $2d$ many measurements to estimate the two sums on the right hand side of Eq. \eqref{eq:process-general}, which both are really probabilistic expectations of random variables distributed according to $\lambda$ or uniformly, respectively. The values of the random variables, of the form $F(\ket{\phi};\mathcal{T})^2$, are quantum-mechanical expectations, which themselves can be interpreted as probabilistic expectations of Bernoulli variables with parameter $F(\ket{\phi};\mathcal{T})^2$. Thus, by sampling $j_i$ independently $n$ times according to $\lambda$, then carrying out the test corresponding to $F(\ket{j_i};\mathcal{T})^2$ (i.e., prepare $\ket{j_i}$, apply $\mathcal{T}$, measure $\proj{j_i}$) and recording the outcome in the variable $X_i\in\{0,1\}$: we can estimate $\sum_j \lambda_j F(\ket{j}^{B};\mathcal{T})^2$ by the empirical mean $\frac{1}{n} \sum_i X_i$ within error $\epsilon$, except with probability $\leq 2 e^{-2n\epsilon^2}$; 
similarly for $\sum_{\hat{\alpha}} \frac{1}{d} F(\ket{\psi_{\hat{\alpha}}}^{B};\mathcal{T})^2$. This holds by Hoeffding's tail inequality \cite{Concentration}. 
In other words, we can estimate the entanglement fidelity of a channel using a number $n$ of pure state samples that only depends on the desired accuracy and confidence, but is independent of the dimension of the quantum system.

\section{Towards multipartite states}
\label{sec:multiparty}
Certifying general multipartite states requires a systematic approach, which we develop here. To set a benchmark for efficiency, it is useful to first recall that a conventional $n$-qubit stabilizer state is uniquely certified by only $n$ local generators from the Pauli group, establishing a remarkable standard of linear scaling. While specific fidelity certification methods like the above exist for certain classes of highly entangled states such as GHZ-like states, and also for non-stabilizer states like W-type and Dicke states, a generally applicable method is still needed \cite[App.~A.6]{Bavaresco:PhD}. 

To start, the ideas from the previous section can be applied recursively: note that for any measurement on one party, we get a projected state of the remaining $n-1$, and the projectors $P$ and $Q$ certify this projected state for each measurement outcome. Thus, we should be able to do that with a one-way LOCC protocol going from one party to the next in a prescribed order, leading to a set of $2^{n-1}$ separable projectors whose unique joint $+1$-eigenstate is $\ket{\psi}$. 

The simplest case is $n=3$, and denote the three parties $A$, $B$ and $C$. We apply Lemma \ref{lemma:little} first to the given tripartite state $\ket{\psi}$ in the bipartite cut $A:BC$, yielding the two projectors
\begin{equation}
  \label{eq:base-case}
  P^{(0)} = \sum_j \proj{j}^A\otimes\proj{j}^{BC},
  \quad
  P^{(1)} = \sum_{\hat{\alpha}} \proj{\hat{\alpha}}^A\otimes\proj{\psi_{\hat{\alpha}}}^{BC}, 
\end{equation}
such that $\proj{\psi} = P^{(0)}P^{(1)}$.
By construction they are separable with respect to the $A:BC$ bipartite cut, but evidently in general not with respect to any other bipartition of the system. However, now we can apply Lemma \ref{lemma:little} to each of the bipartite states $\proj{j}^{BC}$ and $\proj{\psi_{\hat{\alpha}}}^{BC}$ on $B\otimes C$:
\[
  \proj{j}^{BC} = P_jQ_j,
  \quad
  \proj{\psi_{\hat{\alpha}}}^{BC} = P_{\hat{\alpha}} Q_{\hat{\alpha}},
\]
and inserting this into Eq. \eqref{eq:base-case} and using Lemma \ref{lemma:little} once again we find 
\begin{equation}\begin{split}
  \label{eq:induction-step}
  P^{(0)} = P^{(00)}P^{(01)} \text{ with } [P^{(00)},P^{(01)}]=0, \\
  P^{(1)} = P^{(10)}P^{(11)} \text{ with } [P^{(10)},P^{(11)}]=0, 
\end{split}\end{equation}
where
\begin{equation}\begin{split}
  \label{eq:next-case}
  P^{(00)} &= \sum_j \proj{j}^A\otimes P_j^{BC}, \\
  P^{(01)} &= \sum_j \proj{j}^A\otimes Q_j^{BC}, \\
  P^{(10)} &= \sum_{\hat{\alpha}} \proj{\hat{\alpha}}^A\otimes P_{\hat{\alpha}}^{BC}, \\
  P^{(11)} &= \sum_{\hat{\alpha}} \proj{\hat{\alpha}}^A\otimes Q_{\hat{\alpha}}^{BC}.
\end{split}\end{equation}
These four projectors are evidently separable, as indeed are their complements $\1-P^{(ab)}$, since the POVMs $(P^{(ab)},\1-P^{(ab)})$ are all implementable by a one-way LOCC procedure starting at $A$, passing to $B$ and ending at $C$. 
However, unlike the Pauli stabilizers of multipartite stabilizer states, they do not seem to commute pairwise, only in certain combinations:
\[
  [P^{(00)},P^{(01)}] = 
  [P^{(10)},P^{(11)}] =  
  [P^{(0)},P^{(1)}] = 0.
\]

Still, this is enough to get a tripartite generalisation of Theorem \ref{theorem:fidelity}: 
\begin{theorem}
  \label{theorem:fidelity-operator-3}
  For any pure state $\proj{\psi}$ in the tripartite system $A\otimes B\otimes C$ and the projectors $P^{(ab)}$, defined above for $a,b\in\{0,1\}$,
  \begin{equation}\begin{split}
    \label{eq:fidelity-operator-3}
    \proj{\psi} 
     &= P^{(00)}P^{(01)}\, P^{(10)}P^{(11)}, \\
    \1-\proj{\psi} 
     &\leq \sum_{a,b=0}^1 \left( \1-P^{(ab)} \right).
  \end{split}\end{equation}
  Consequently, for any state $\rho$ in the system $A\otimes B\otimes C$,
  \[
    F\!\left(\rho,\psi\right)^2 \!= \tr\rho\psi \geq \tr\rho P^{(00)} + \tr\rho P^{(01)} + \tr\rho P^{(10)} + \tr\rho P^{(11)} - 3. 
  \]
\end{theorem}
\begin{myproof}
The stabilizer equation follows by multiplying out the product on the right hand side by commuting pairs:
\[
  P^{(00)}P^{(01)}\, P^{(10)}P^{(11)} 
   = P^{(0)}P^{(1)}
   = \proj{\psi}.
\]

To generalise it to the operator inequality, starting with the reasoning of Theorem \ref{theorem:fidelity}, we have 
\[
  \1-\psi \leq \1-P^{(0)} + \1-P^{(1)}. 
\]
Now, we can use $P^{(a)} = P^{(a0)}P^{(a1)}$ and $[P^{(a0)},P^{(a1)}]=0$ for $a=0,1$ to bound in turn
\[
  \1-P^{(a)} \leq \1-P^{(a0)} + \1-P^{(a1)},
\]
and inserting the latter inequalities into the former we obtain the claim.
\end{myproof}

\medskip
Straightforward iteration on of this prescription on an $n$-party state yields projectors $P^{(\underline{u})}$ for binary sequences (words) $\underline{u}=u_1\ldots u_k \in\{0,1\}^k$, $0\leq k < n$, where we denote the empty word of length $0$ by $\emptyset$ and $P^{(\emptyset)} = \proj{\psi}$, with recursive commutation and stabilizer properties as follows. 
\begin{theorem}
  \label{thm:fidelity-n}
  For any $n$-party pure state $\proj{\psi}$, the projectors $P^{(\underline{u})}$ and $\1-P^{(\underline{u})}$ are separable for all words $\underline{u}$ of length less than $n$, and
  \begin{equation}\begin{split}
    \label{eq:fidelity-operator-n}
    \proj{\psi} 
     &= \prod_{\underline{u}\in\{0,1\}^{n-1}} P^{(\underline{u})}, \\
    \1-\proj{\psi} 
     &\leq \sum_{\underline{u}\in\{0,1\}^{n-1}} \left( \1-P^{(\underline{u})} \right), 
  \end{split}\end{equation}
  where the product in the first equation is intended in lexicographic order.
  
  More precisely, $P^{(\emptyset)} = \proj{\psi}$ and for any word $\underline{u} \in \{0,1\}^k$ of length $k<n$,
  \[
    P^{(\underline{u})} = P^{(\underline{u}0)}P^{(\underline{u}1)}, 
    \text{ and }
    [P^{(\underline{u}0)},P^{(\underline{u}1)}]=0.
  \]
\end{theorem}
\begin{myproof}
The claim follows directly from the preceding arguments by induction on the number $n$ of parties.
\end{myproof}

\medskip
Theorems \ref{theorem:fidelity}, \ref{theorem:fidelity-operator-3} and \ref{thm:fidelity-n} realise -- in principle -- what we have set out to achieve: to certify any given multipartite pure state with a finite number of separable stabilizer projections. Furthermore, each of these projections is implemented as part of a one-way LOCC protocol (by which the parties are arranged in an arbitrary pre-agreed order), and the infidelity with the target state is bounded in terms of the failure probability of passing the quantum tests with the stabilizer projections. The latter is important to be able to determine error bars under unsharp measurements and finite statistics. 

However, there are important differences to the Pauli stabilizer formalism. The first is that there we only need a number of projections equal to the number $n$ of parties, whereas here we have an exponential scaling of required projections ($2^{n-1}$). Secondly, the number of local measurements needed is different, namely up to $d^2$ stabilizer bases at each of the qudits for Pauli stabilizer states, but exponentially increasing numbers of measurements depending on the position in the ordering of the parties. To wit, the first party requires two complete projective measurements, the second $4d$, the third $8d^2$, \ldots, the $k$-th $2(2d)^{k-1}$, except for the last ($n$-th), which ``only'' has $(2d)^{n-1}$ binary measurements. In other words, from the third party in the chain onwards the scheme requires more different local measurements than full tomography.

\section{Discussion and open questions}
\label{sec:open}
We have presented a construction of separable (in the multipartite case: fully separable) projectors suitable for local and robust certification of pure entangled states. As a matter of fact, it describes the eigenprojectors of measurements implementable by one-way LOCC (in the multipartite case this entails an order in which the quantum parties act one after the other, and broadcast classical communication to subsequent parties). 

Our approach differs fundamentally from general-purpose methods like classical shadow tomography \cite{Aaronson}, which use randomized local measurements to construct an approximation of the observable $\1-\proj{\psi}$. While this presents itself as a general-purpose recipe, its efficiency is not necessarily optimal for the specific task of certifying fidelity to a particular pure state. 
The method in \cite{HuangPreskillSoleimanifar} is inspired by shadow tomography, but except for one party all others perform only computational-basis measurements. Because of this, it does not work for a submanifold of states (which however includes EPR and GHZ states), and when it does may incur an arbitrarily large inefficiency overhead. Their approach was refined in \cite{GuptaHeODonnell}, using adaptivity, and resulting in a method that can efficiently certify any state sufficiently close to the target state.
Instead, we construct a small set of projectors tailored to the specific structure of the target state via its Schmidt decomposition. This distinction is crucial, as our approach may offer a more direct and resource-efficient path for the specific task of fidelity estimation with high confidence, as opposed to general state characterization/certification.
This feature distinguishes it, too, from information-based approaches such as \cite{Huang-et-al}, which inherently can only work for highly entangled states within the given dimension.

A key feature of this bespoke construction is the structural relation $PQ = \proj{\psi}$, which is a stronger condition than merely having $\ket{\psi}$ as the unique common $+1$-eigenstate (which is the basis of the approach in \cite{YuShangGuehne-1}). This relation implies that operator sequences such as $PQP$ also function as exact projectors onto the target state, since $PQP = P(QP) = P\proj{\psi} = \proj{\psi}$. This mechanism ensures a complete suppression of the orthogonal subspace, where the sequential application of the projectors annihilates any state component orthogonal to $\ket{\psi}$. It remains to be determined whether such a strong structural relation between the certifying projectors is a necessary feature for robust, local certification, or if it is a specific artifact of our construction. It does give us the matrix upper bound on $\1-\proj{\psi}$ in Eqs.~\eqref{eq:fidelity-operator}, \eqref{eq:fidelity-operator-3} and \eqref{eq:fidelity-operator-n}, which translate into statistical guarantees when estimating the target state fidelity in terms of the stabiliser projectors. On the other hand, however, \cite{YuShangGuehne-1} and \cite{YuShangGuehne-2} offer a direct analysis of the best possible statistical performance in terms of the eigenvalue gap of a POVM element constructed from the available test measurements. The use of randomised bases links these works with \cite{Ohad-et-al}. Note that those works, by maximising the eigenvalue gap, may incur a higher complexity of the procedure in terms of the number of LOCC measurements involved, which in fact scales with the dimension.

The present work raises several questions:

\begin{enumerate}
\item Can we get a better scaling of the number of projectors with the number of parties? Note that Pauli stabilizer states require $n$ pairwise commuting projectors to identify their unique stabilizer state, rather than the $2^{n-1}$ we needed in Theorem \ref{thm:fidelity-n}. Can we get a $O(n)$ scaling, or at least a polynomial in $n$? This would be important for two practical reasons: first, it would decrease the number of measurements required, and secondly the build-up of errors from the estimation of each observable in Eq. \eqref{eq:fidelity-operator-n} would be reduced from $2^{n-1}\epsilon$ to a smaller multiple of $\epsilon$. 


\item What does it change for Theorem \ref{theorem:fidelity} if we only want that $\ket{\psi}$ is the unique joint $+1$-eigenstate of the separable POVM elements $P$ and $Q$, but say with a bound on the second-largest eigenvalue of $PQP$? Note that the ultimate target is an inequality in the vein of Eq.~\eqref{eq:fidelity-operator} [and similarly Eqs.~\eqref{eq:fidelity-operator-3} and \eqref{eq:fidelity-operator-n}], relating the infidelity to the probability of failing the LOCC tests corresponding to the separable stabilizers. 

\item Are there experimental applications of our multipartite stabilizers, perhaps to certify high-dimensional entanglement beyond bipartite systems, generalising \cite{Bavaresco-et-al:fid}?

\item For Theorems \ref{theorem:fidelity},  \ref{theorem:fidelity-operator-3} and \ref{thm:fidelity-n}, the ultimately important fact is the fidelity bound (operator inequality), and not so much the pairwise commutativity of the projectors (which however makes the construction more similar to Pauli stabilizers). However, this motivates a rather more mathematical question: is it even possible to obtain $\ket{\psi}$ as the unique $+1$-eigenstate of some number of commuting separable projectors? Theorem \ref{theorem:fidelity} does that for $n=2$ with the minimum number two of projectors. There is however an elegant proof of principle of this based on the fact that $\1-\proj{\phi}$ is separable for any bipartite $\ket{\phi}\in A\otimes B$. This is true because that operator is in the Hilbert-Schmidt (aka Frobenius, aka Schatten-$2$) norm ball of radius $1$ around $\1$, which consists entirely of separable states \cite{GurvitsBarnum-1}. Complete $\ket{\psi}=:\ket{\phi_1}$ to an orthonormal basis of $A\otimes B$ of vectors $\ket{\phi_j}$, and let  $P_j=\1-\proj{\phi_j}$, $j=2,\ldots,|A||B|$; this does the job, as $\proj{\psi} = \prod_{j=2}^{|A||B|} P_j$. 
Can we emulate this for $n>2$ parties, i.e.~is $\1-\proj{\phi}$ fully separable for any pure state $\ket{\phi}$ in any number of parties? Unfortunately, the fully separable balls around $\1$ for $n>2$ parties have Hilbert-Schmidt radius less than $1$ and they decrease exponentially with $n$ \cite{AubrunSzarek:sep,GurvitsBarnum-2,GurvitsBarnum-3}. Note however that by \cite{GurvitsBarnum-1}, $\1-\proj{\phi}$ is biseparable simultaneously with respect to every bipartite cut of the $n$ systems.
\end{enumerate}

\section*{Acknowledgments}
The authors wish to thank Marcus Huber and Jessica Bavaresco for discussions in the wake of the latter's doctoral thesis defence and about the complexity of certifying entanglement by local measurements in general. Furthermore, we thank William Rosewood and John Taggart for their insistence on obtaining results by regular means.
We also thank Otfried G\"uhne for bringing the references \cite{YuShangGuehne-1} and \cite{YuShangGuehne-2} to our attention.
JA and AW are supported by the Spanish MICIN with funding from European Union NextGenerationEU (PRTR-C17.I1) and the Generalitat de Catalunya, and by the Spanish MICIN (project PID2022-141283NB-I00) with the support of FEDER funds.
AW was or is furthermore supported by the European Commission QuantERA grant ExTRaQT (Spanish MICIN project PCI2022-132965), by the Spanish MTDFP through the QUANTUM ENIA project: Quantum Spain, funded by the European Union NextGenerationEU within the framework of the ``Digital Spain 2026 Agenda'', by the Alexander von Humboldt Foundation, and by the Institute for Advanced Study of the Technical University of Munich.


\begin{thebibliography}{20}
\bibitem{Aaronson} Scott Aaronson.
``Shadow tomography of quantum states'', in:
\emph{Proc. 50th Annual ACM SIGACT Symposium on Theory of Computing}, pp. 325-338 (2018).

\bibitem{Artiles} Luis M. Artiles, Richard D. Gill, and M\u{a}d\u{a}lin I. Gut\u{a}.
``An invitation to quantum tomography'', \emph{Journal of the Royal Statistical Society Series B: Statistical Methodology} {\bf 67}(1):109-134 (2005); arXiv:math/0405595v2.

\bibitem{AshikhminKnill} Alexei Ashikhmin and Emanuel Knill. 
``Nonbinary quantum stabilizer codes'',
\emph{IEEE Transactions on Information Theory} {\bf 47}:3065–3072 (2001).

\bibitem{AubrunSzarek:sep} Guillaume Aubrun and Stanis{\l}aw Szarek. 
``Tensor products of convex sets and the volume of separable states on $N$ qudits'', \emph{Physical Review A} {\bf 73}:022109 (2006); arXiv:quant-ph/0503221.

\bibitem{Bavaresco-et-al:fid} Jessica Bavaresco, Natalia Herrera Valencia, Claude Kl\"ockl, Matej Pivoluska, Paul Erker, Nicolai Friis, Mehul Malik, and Marcus Huber.
``Measurements in two bases are sufficient for certifying high-dimensional entanglement'', \emph{Nature Physics} {\bf 14}:1032 (2018); arXiv[quant-ph]:1709.07344.

\bibitem{Bavaresco:PhD} Jessica Bavaresco.
\emph{Certifying complex quantum properties: High-dimensional entanglement and indefinite causal order}, PhD thesis, Universt\"at Wien, Department of Physics, 2021.

\bibitem{Bennett:Tele} Charles H. Bennett, Gilles Brassard, Claude Cr{\'e}peau, Richard Jozsa, Asher Peres and William K. Wootters.
``Teleporting an unknown quantum state via dual classical and Einstein-Podolsky-Rosen channels'', \emph{Physical Review Letters} {\bf 70}(13):1895-1900 (1993).

\bibitem{BertlandKrammer:fullTom} Reinhold A. Bertlmann and Philipp Krammer.
``Bloch vectors for qudits'', \emph{Journal of Physics A: Mathematical and Theoretical} {\bf 41}:235303 (2008); arXiv[quant-ph]:0806.1174.

\bibitem{Concentration} St\'ephane Boucheron, G\'abor Lugosi, and Olivier Bousquet. ``Concentration Inequalities'', In: Olivier Bousquet, Ulrike von Luxburg, and Gunnar R\"atsch (eds.), \emph{Advanced Lectures on Machine Learning. ML 2003}, Lecture Notes in Computer Science (LNAI), vol. 3176, pp. 208-240, Springer Verlag, Berlin, Heidelberg, Accra, 2004; DOI: 10.1007/978-3-540-28650-9{\_}9.

\bibitem{LOCC} Eric Chitambar, Debbie Leung, Laura Man\v{c}inska, Maris Ozols, and Andreas Winter. 
``Everything You Always Wanted to Know About LOCC (But Were Afraid to Ask)'', \emph{Communications in Mathematical Physics} {\bf 328}(1):303-326 (2014); arXiv[quant-ph]:1210.4583. 

\bibitem{Ekert1991} Artur K. Ekert.
``Quantum cryptography based on Bell’s theorem'', \emph{Physical Review Letters} {\bf 67}(6):661 (1991). 

\bibitem{Gottesman:PhD} Daniel Gottesman. 
\emph{Stabilizer Codes and Quantum Error Correction}, PhD thesis, Caltech, 1997;
arXiv:quant-ph/9705052.

\bibitem{GuptaHeODonnell} Meghal Gupta, William He, and Ryan O'Donnell. ``Few Single-Qubit Measurements Suffice to Certify Any Quantum State'', arXiv[quant-ph]:2506.11355 (2025).

\bibitem{GurvitsBarnum-1} Leonid Gurvits and Howard N. Barnum. 
``Largest separable balls around the maximally mixed bipartite quantum state'', \emph{Physical Review A} {\bf 66}:062311 (2002); arXiv:quant-ph/0204159.

\bibitem{GurvitsBarnum-2} Leonid Gurvits and Howard N. Barnum. 
``Separable balls around the maximally mixed multipartite quantum states'', \emph{Physical Review A} {\bf 68}:042312 (2003); arXiv:quant-ph/0302102.

\bibitem{GurvitsBarnum-3} Leonid Gurvits and Howard N. Barnum. 
``Better bound on the exponent of the radius of the multipartite separable ball
'', \emph{Physical Review A} {\bf 72}:032322 (2005); arXiv:quant-ph/0409095.

\bibitem{Hofmann:fid} Holger F. Hofmann.
``Complementary classical fidelities as an efficient criterion for the evaluation of experimentally realized quantum operations'',
\emph{Physical Review Letters}, {\bf 94}:160504 (2005); arXiv:quant-ph/0409083.

\bibitem{Horodeckix4:review} Ryszard Horodecki, Pawe{\l} Horodecki, Micha{\l} Horodecki, and Karol Horodecki.
``Quantum entanglement'', \emph{Reviews of Modern Physics} {\bf 81}(2):865-942 (2009); arXiv:quant-ph/0702225v2.

\bibitem{Hostens-et-al} Erik Hostens, Jeroen Dehaene, and Bart De Moor.
``Stabilizer states and Clifford operations for systems of arbitrary dimensions, and modular arithmetic'', \emph{Physical Review A} {\bf 71}:042315 (2005); arXiv:quant-ph/0408190.

\bibitem{HuangPreskillSoleimanifar} Hsin-Yuan Huang, John Preskill, and Mehdi Soleimanifar. 
``Certifying almost all quantum states with few single-qubit measurements'', in: \emph{Proc. 2024 IEEE 65th Annual Symposium on Foundations of Computer Science (FOCS)}, pp. 1202-1206 (2024); arXiv[quant-ph]:2404.07281. 

\bibitem{Huang-et-al} Zixin Huang, Lorenzo Maccone, Akib Karim, Chiara Macchiavello, Robert J. Chapman, and Alberto Peruzzo.
``High-dimensional entanglement certification'', \emph{Scientific Reports} {\bf 6}(1):27637 (2016); arXiv[quant-ph]:1604.05824.

\bibitem{KKKS} Avanti Ketkar, Andreas Klappenecker, Santosh Kumar, and Pradeep Kiran Sarvepalli.
``Nonbinary Stabilizer Codes Over Finite Fields'', \emph{IEEE Transactions on Information Theory} {\bf 52}(11):4892-4914 (2006). 

\bibitem{Ohad-et-al} Ohad Lib, Shuheng Liu, Ronen Shekel, Qiongyi He, Marcus Huber, Yaron Bromberg, and Giuseppe Vitagliano.
``Experimental certification of high-dimensional entanglement with randomized measurements'', \emph{Physical Review Letters} {\bf 134}:210202 (2025); arXiv[quant-ph]:2412.04643.

\bibitem{Nielsen} Michael A. Nielsen.
``Conditions for a Class of Entanglement Transformations'', \emph{Physical Review A} {\bf 83}(2):436 (1999); arXiv:quant-ph/9811053.

\bibitem{NielsenAndChuang}  Michael A. Nielsen, and Isaac L. Chuang.
\emph{Quantum Computation and Quantum Information}, Cambridge University Press, Cambridge, 2010.

\bibitem{TadejZyczkowski:Hadamard} Wojciech Tadej and Karol \.{Z}yczkowski. 
``A Concise Guide to Complex Hadamard Matrices'',
\emph{Open Systems and Information Dynamics} {\bf 13}(2):133-177 (2006); arXiv:quant-ph/0512154.

\bibitem{Werner} Reinhard F. Werner. 
``Quantum states with Einstein-Podolsky-Rosen correlations admitting a hidden-variable model'',
\emph{Physical Review A} {\bf 40}(8):4277 (1989).

\bibitem{YuShangGuehne-1} Xiao-Dong Yu, Jiangwei Shang, and Otfried G\"uhne. ``Optimal verification of general bipartite pure states'', \emph{npj Quantum Information} {\bf 5}:112 (2019); arXiv[quant-ph]:1901.09856.

\bibitem{YuShangGuehne-2} Xiao-Dong Yu, Jiangwei Shang, and Otfried G\"uhne. ``Statistical Methods for Quantum State Verification and Fidelity Estimation'', \emph{Advanced Quantum Technologies} {\bf 5}:2100126 (2022); arXiv[quant-ph]:2109.10805.

\end{thebibliography}
\end{document}